# High-temperature PbTe diodes

Z. Dashevsky*, V. Kasiyan *, E. Mogilko**, A.Butenko**, R. Kahatabi**, S. Genikov*, V.Sandomirsky**, Y.Schlesinger**

*Materials Engineering Department, Ben-Gurion University of the Negev, P.O.B. 653, Beer-Sheva 84105, Israel*

**Department of Physics, Bar-Ilan University, Ramat-Gan 52900, Israel*

**Abstract**

We describe the preparation of high-temperature PbTe diodes. Satisfactory rectification was observed up to 180-200 K. Two types of diodes, based on a *p*-PbTe single crystal, were prepared: (1) by In ion-implantation, and (2) by thermodiffusion of In. Measurements were carried-out from ~ 10 K to ~ 200 K. The ion-implanted diodes exhibit a satisfactorily low saturation current up to a reverse bias of ~ 400 mV, and the thermally diffused junctions up to ~ 1 V. The junctions are linearly graded. The current-voltage characteristics have been fitted using the Shockley model. Photosensor parameters: zero-bias-resistance × area product, the $R_0C$ time-constant and the detectivity $D^*$ are presented.

**1. Introduction**

The band gap of PbTe corresponds to the IR spectral region (2 – 6 μm), thus, the PbTe *p-n* junction is often related to as a prospective photovoltaic element [1].

We describe here the preparation of two types of junctions, differing in the method used to create the *n*-region on the original *p*-type PbTe ingot: one by $In_4Te_3$ gas diffusion and the other by Indium ion-implantation.

In the following we present the details of the preparation methodology used, and the characterization of these diodes by capacitance vs. voltage (*C-V*) and current vs. voltage (*I-V*) characteristics measurements, over a wide temperature interval. We derived the parameters essential for evaluation of the photosensor qualities: the dynamical resistance at zero bias, $R_0(T)$, the zero-bias-resistance × area product, $R_0(T)A$, the saturation current, $I_0(T)$, the $R_0C$-time constant (*C* is the junction capacitance) and the estimated detectivity, $D^*$.

## 2. Sample preparation.

### *2.1. The crystal growth*

PbTe crystals were grown by the Czochralski technique [2]. In the crystal growing process, polycrystalline PbTe is placed in the crucible and the furnace is heated to above the melting temperature of PbTe. A <100> oriented seed PbTe crystal was used. To prevent decomposition of the melt, liquid encapsulation by, about 1 cm thick, molten boron trioxide ($B_2O_3$) layer, was employed. Boron trioxide is inert to PbTe at the grow process temperature. The temperature gradient at the crystallization front and the crystal pull rate were of the order of 20–25 $Kcm^{-1}$ and 5-10 mm/h, respectively. To enhance the equalization of the heat flux through the lateral surface of the crystal, the crucible and the crystal were rotated, in opposite directions, at an angular rate of 1 $s^{-1}$. The crystal orientation was determined by the Laue diffraction. Back reflection photographs of an <100> oriented PbTe crystal clearly show the four-fold symmetry and the absence of any other reflections.

The crystals were grown with an excess of Te (up to 1 at %). Deviations from the stoichiometric composition at elevated temperatures induce the formation of lattice defects, vacancies and possibly interstitials. These defects can be quenched-in by sufficiently rapid cooling, to inhibit their contribution to the electrical conduction.

The Hall effect is employed for direct measurement of the carrier concentration. At low temperature (80 K), all acceptor impurities are ionized. Under these circumstances the carrier concentration is equal to the concentration of the lattice point defects, namely, to the excess Te, of the order of $(1\text{-}5) \times 10^{18}$ cm$^{-3}$.

### 2.2. Thermally diffused p-n junction (TDJ)

*Mechanical and electrochemical polishing.* Rectangular wafers (10×20×1 mm) have been cut from the single crystal ingot. One side of these wafers has been successively polished mechanically and electrochemically. For mechanical polish aluminum oxide powders, with grain size: 22, 7, 5, 0.3 and 0.05 μm, were used. We found that the most important condition for the final polish quality is the uniformity of the roughness after the mechanical polish, rather than the value of roughness itself. Thus, eventually, only the first three types of powders, yielding satisfactorily uniform surface roughness, were used. Subsequently, the PbTe plates were polished electrochemically. This process smoothes the mechanically treated surface and relaxes the surface strains caused by it. In the electrochemical bath, the PbTe plate was used as the anode and a Pt plate was used as the cathode. 180 ml of Norr solution electrolyte [3] has been used in the process. The electrochemical treatment consists of two competing processes, oxidation and dissolution of oxide [4, 5]. The dominant result depends on the content of glycerol in the electrolyte. An anodization circuit with constant voltage [5] of 20 V was used for different time durations, ranging from 5 to 30 minutes.

During the electrochemical process, the solution was stirred at a constant rate of about 60 rpm. According to AFM analysis the shortest time, for obtaining satisfactory surface smoothness of ~ 4 nm, is 10 minutes. The rate of the etching is approximately 5 μm per minute.

Thermally diffused doping was accomplished by exposing the sample to indium from a gaseous source of $In_4Te_3$, maintaining a constant level of surface concentration of In during the diffusion anneal. The concentration of the In dopant is in the 0.1 – 4 at% range, and can be varied by changing the exposure time and the annealing temperature (600 to 700 °C). The concentration profile of In within the crystal was assumed to be consistent with the solution of the diffusion equation for a constant surface source.

The determination of the spatial variation of the dopant concentration and the location of the *p-n* junction is of utmost importance. The Seebeck coefficient *S* of electrical conductors is a measure of the carrier concentration and, hence, of the dopant distribution. We made use of a novel, recently developed, method for measuring the local Seebeck coefficient along the surface of the doped crystal [6]. In this method a very thin and sharp probe tip, heated to temperature $T_1$, is put in contact with the surface of the isothermal sample maintained at temperature $T_o$. The measuring probe is actually the junction of a Cu-CuNi thermocouple. A second, similar thermocouple, in good electrical and thermal contact with the sample, its junction temperature being, thus, $T_o$. The voltages, $V_1$ and $V_2$, between the Cu-Cu and the CuNi-CuNi wires, respectively, are measured simultaneously. This device allows a spatial resolution of 10 μm. The Seebeck coefficient variations were measured along a direction normal to the external free surface on a polished plane of the crystal. The transition from *n*-type to *p*-type conductivity (*p-n* junction) causes a change of sign of the Seebeck coefficient and its location can be determined, as described above, by scanning the local Seebeck coefficient. These measurements allowed to determine the diffusion coefficient of In in PbTe (extrapolated to infinite temperature [6]), $D_0 \approx 0.1$ cm$^2$/s, and an activation energy of $\approx 1.4$ eV.

*<u>Contacts and mesa-stucture.</u>* The electrical contacts on both sides of the sample were formed by vapor deposition of a thin (10 nm) layer of Cr followed by deposition of a 100 nm thick layer of Au. The Cr layer ensures a good adhesion of the gold layer to the PbTe surface. The polished, front, side with the Cr and Au point contacts, was then slotted by a diamond saw to produce the mesa array with columns of (0.1-0.5) x (0.1-0.5) mm with 0.2 mm distance between them.

### *2.3. Implanted p-n junctions* (**IJ**) *- The planar array*

The method of ion implantation is well known [7].

<u>*Generation of the native oxide layer*</u>**.** The process of electrochemical etching of PbTe in Norr solution [3, 4] includes two competing processes: development of a native oxide layer and its dissolution. The 1μm thick dielectric oxide layer is formed by electrochemical oxidation of the polished PbTe surface. To create the oxide layer, Norr electrolyte solution was used [3], similar to that used in the electrochemical etching process, except for the glycerol quantity, which was now 4 times larger. The same anodization circuit with constant voltage [5] of 20 V was applied as before. We found that about 17 minutes of the oxidation process is suffice to form an oxide layer of ~ 1μm thickness which was analyzed by Auger spectroscopy. This oxide layer has demonstrated a high breakdown strength.

<u>*Photolitography,*</u> The Norr solution is a strong alkaline (pH > 13). We used two sorts of negative photoresist (PR): SU-8 2000 and SC 100. The most important factors are the gradually increasing spin speed to obtain a uniform spreading of 2 μm thick PR layer, and allowing sufficient time for relaxation after each step of photolithography. Carl-Zeiss MJB3 Mask Aligner was used to transfer the masks patterns onto the substrates, coated with PR, by near-UV (350-400 nm) illumination. After developing, hard bake was made to further cross-link the material on the areas surrounding the future *p-n* junctions. The hard resist is left as a part of the final device, producing further protection of the oxide layer and extra protection against radiation.

To remove the oxide layer in the array, after photolithography, electrochemical etching, using the Norr electrolyte at 20 V, was used. We found that 10 minutes etching was sufficient for a complete removal of the oxide, without damaging the PR layer and, consequently, the underlying oxide layer.

*__Ion implantation of In.__* The samples were exposed to an irradiation dose of up to $10^{16}$ cm$^{-2}$, during 3-4 hours. This allowed the intrinsic diffusion of the irradiation-created vacancies and a homogeneous electron concentration through the whole thickness of the sample (up to 4 μm) [7]. Thus, an *n*-type layer, with an electron concentration of $5\times10^{18}$ cm$^{-3}$, was formed up to a depth of 3-5 μm from the surface of the original *p*-type PbTe sample.

*__Contacts arrangement.__* Cr and Au were vacuum deposited through the masks into the array "windows" and on the back side of the sample, for the electrical contacts. The electric contacts were fitted with In and thin Au wires.

**3. Characterization of *p-n* junctions**

The schematic drawing of the two types of diodes, is presented in Fig.1. The acceptor concentration of the as-grown *p*-PbTe crystal is $N_a \approx 5\times10^{17}$ cm$^{-3}$. The diffusion length was ≈ 30 μm. The mobility is ≈ 14000 cm$^2$/V·s at 80 K. The concentration of donors, introduced by diffusion or implantation, was $N_d \approx 5\times10^{18}$ cm$^{-3}$. The junction in implanted diodes was located at depth of several microns below the PbTe-surface, and at about 70 μm in the thermally diffused junctions.

The *C-V* and *I-V* curves were measured over a temperature interval 12 – 200 K. The investigated samples were placed in a He gas closed-cycle refrigerator cryostat, in a vacuum of about $10^{-7}$ Torr. The *I-V* characteristics were obtained using a 2410 Keithley SourceMeter. The applied bias voltage varied between –1 V and +0.5 V, measurements being taken every 0.2 mV. The positive electrode was connected to the *p* side of the diode.

A QuadTech 1920 LCR was used to measure the *C-V* characteristics. The LCR meter, is operating using the 4- point method. The capacitance was measured using an alternating signal of 20 mV at a frequency of 1 MHz, with the applied reverse bias voltage varying between 0 V and -1 V, measurements being taken every 5 mV.

*Capacitance-Voltage characteristics.* The *C-V* curves were measured as a function of temperature and of bias voltage. The *C(V)* data were analyzed by the standard theory [8]. The linearity of the $C^{-3}$ vs. *V* curves clearly indicates that both types of junctions are linearly graded.

The dielectric constant $\varepsilon$ of the TDJ and In-IJ diodes, extracted from the measured *C(V)* curves, is shown in Fig. 2b. The depletion layer width *W* and the dopant concentration at the depletion region edge, $N_D$, vary from 0.9 µm and $0.9 \times 10^{17}$ cm$^{-3}$, respectively, at 15 K, to 0.6 µm and $0.6 \times 10^{17}$ cm$^{-3}$, at 170 K, for TDJ, while the corresponding values for IJ are: 0.15 µm and $35 \times 10^{17}$ cm$^{-3}$ at 15 K, and 0.1 µm and $25 \times 10^{17}$ cm$^{-3}$ at 120 K.

It should be noted that $\varepsilon$ varies strongly with temperature, a fact that has been frequently overlooked in past works.

The gradient of the dopant concentration *a* in the case of In-IJ, ~ $4.5 \times 10^{31}$ m$^{-4}$, is larger then that of the TDJ diode, ~ $2 \times 10^{29}$ m$^{-4}$, by more than two orders of magnitude. Accordingly, the width of the depletion region in the In-IJ, ~ 0.1 µm, is smaller by about a factor of 7 than that of the TDJ diode ~ 0.7 µm, and the dopant concentration in In-IJ at *W/2* is markedly higher than in the TDJ diode.

*Current-Voltage characteristics.* The experimental *I-V* curves have been fitted by the Shockley formula

$$I = I_0 \left[ \exp\left(\frac{eV}{n \cdot kT}\right) - 1 \right], \tag{1}$$

where $I_0$ is the saturation current and $n$ is the ideality factor, $k$ is the Boltzmann constant. Typical examples of the fitting for the TDJ diodes are shown in Fig. 3. The shape of the *I-V* curve and the quality of the fit are similar to those of In-IJ diodes.

The plot of log $R_0$ vs. $1/T$ ($R_0 = R_{V=0}$) is presented in Fig. 4. We observe two temperature regions with different activation energies $E_a$ for both types of diodes. A high-temperature region ($T > 50$ K) with $E_a = 105$ meV, and a low temperature region ($T \lesssim 50$ K) with $E_a = 0.22 – 0.3$ meV.

The temperature dependence of the derived ideality factor $n$ and of the saturation current density $j_0$ are shown in Figs 5a and 5b. The $j_0$ vs. $1/T$ plot shows again evidence of two temperature regions with different activation energies. Both types of diodes exhibit a high-temperature activation energy of the saturation current of ~ 135 -145 meV. These features are here reported for the first time.

There are two main contribution to the diode current [8] – the diffusion current ($J_d$) and recombination current ($J_r$). According to the theory $\ln J_d \propto E_g(0\ K)$, and $\ln J_r \propto E_g(0\ K)/2$ ($E_g$ is the band gap). Thus, the recombination current contribution is essential. The value of the ideality coefficient of the TDJ in the high-temperature region, $n \approx 2$, points also to this fact. However, for the In-IJ diode the value of $n$ is much larger than 2 in a sizeable part of the high-temperature region.

The, small activation energy at low temperatures, is connected, conceivably, with a tunneling process. As the direct "band-to-band" tunneling is highly improbable for a broad depletion region, we suggest that the tunneling proceeds via intermediate localized states in the gap.

***Photosensitivity characteristics.*** The zero-bias-resistance × area product, $R_0A$, values obtained, allow to estimate the specific detectivity ($D^*$) of these diodes.

The specific detectivity of a photodiode is defined as [9]

$$D_\lambda^* = \frac{R_\lambda}{\left(\frac{4kT}{R_0 A} + 2e^2\eta Q_B\right)^{1/2}}; \quad R_\lambda = \eta e\lambda/hc; \quad Q_B(\nu_c, T) = \int_{\nu_c}^{\infty} J(\nu)\,d\nu$$

$$J(\nu) = \frac{8\pi\nu^2}{c^2} \cdot \frac{1}{\exp(h\nu/kT) - 1}$$

(2)

Here, $R_\lambda$ is the current mode responsivity at a wavelength $\lambda$; $\eta$ is the quantum efficiency, $\nu_c$ or $\lambda_c$ are the cut-off frequency or wavelength, respectively. The values used for our estimates (assuming black body radiation at 300 K) were $\eta \approx 0.5$, and $R_\lambda$ ($\lambda_c$=4μm) $\approx$ 1.6 A/W. $J(\nu)$ is the flux density per unit frequency interval. $R_0 \cdot A$ is the zero-bias-resistance × area product. The values of some of the photosensitivity parameters, at a number of temperatures, are given in the Table. The detectivity includes two contributions. The first term in the denominator is Johnson noise while the second is the background-induced shot noise. The values of $D^*$ at 12.6 K and 80 K are determined by the background radiation only. The column of $D^*_J$ is the detectivity limited only by the Johnson noise.

Zogg et al [10] have made significant progress in the preparation of high-quality PbTe *p-n* junctions (see also Ref. [1]). They characterized the epitaxial PbTe *p-n* diodes grown on Si substrate with $CaF_2$ buffer layer. Their value of $R_0A$(80 K) $\approx$ 20 Ω·cm² (Fig. 4 in Ref. [10]) is more than an order of magnitude smaller than the values given in the Table.

Recently, Barros et al [11] have characterized the performance of PbTe *p-n* junction, at 80 K, prepared on $BaF_2$. Their values of $R_0A$ (0.23 – 31.8 Ω·cm²) and $D^*$ (5.96×10⁸ - 8.01×10¹⁰ cm·Hz$^{1/2}$/W) obtained in different samples, are markedly lower than our results.

**4. Summary**

Thermally diffused and ion-implanted PbTe *p-n* junction diodes were prepared. Their properties have been thoroughly investigated and characterized over a wide region of temperature and bias voltages. From an analysis of the *C-V* and *I-V* measurements, all the physical parameters of the diodes were derived. The temperature dependence of the static dielectric constant of PbTe was determined, for the first time, from 10 K up to 300 K. The values of the measured and of the estimated parameters of these diodes demonstrate their high photosensor qualities. It should be stressed that the high operating temperature, up to 200 K, of the photodetectors described in this work, obviates the need for cryogenic cooling.

**Figure captions.**

Fig. 1  Schematic drawing of the sample preparation configuration; (a) IJ and (b) TDJ

Fig. 2  The plot of: (a) the inverse cube of the capacitance vs. bias voltage, and

(b) the temperature dependence of the static dielectric constant.

Fig. 3  Fitting of the measured I-V characteristics using Shockley formula, at

(a) 80 K, (b) 130 K, (c) 200 K.

Fig. 4  The temperature dependence of the zer-bias resistance.

Fig. 5  The temperature dependence of: (a) the ideality factor $n$, and (b) the saturation current density $j_0$.

Table 1    Representative values of the photosensitivity parameters of the two types of diodes, IJ and TDJ, at several temperatures.

| Sample | $T$ K | $R_0A$ Ohm cm$^2$ | $R_0C_0$ sec | $D^*$ cm·Hz$^{1/2}$/W | $D_J^*$ cm·Hz$^{1/2}$/W |
|---|---|---|---|---|---|
| IJ | 180 | 0.12 | - | $5.6 \cdot 10^9$ | $5.6 \cdot 10^9$ |
| IJ | 80 | 250 | $1.82 \cdot 10^{-3}$ | $1.2 \cdot 10^{11}$ | $3.8 \cdot 10^{11}$ |
| IJ | 12.6 | $1.55 \cdot 10^4$ | 0.17 | $1.2 \cdot 10^{11}$ | $7.6 \cdot 10^{12}$ |
| TDJ | 180 | 0.38 | $3.25 \cdot 10^{-7}$ | $9.95 \cdot 10^9$ | $9.95 \cdot 10^9$ |
| TDJ | 80 | 845 | $9.06 \cdot 10^{-4}$ | $1.2 \cdot 10^{11}$ | $7.1 \cdot 10^{11}$ |
| TDJ | 12.7 | $8.3 \cdot 10^4$ | 0.086 | $1.2 \cdot 10^{11}$ | $1.7 \cdot 10^{13}$ |

$A$ – sample area = 0.64 mm$^2$

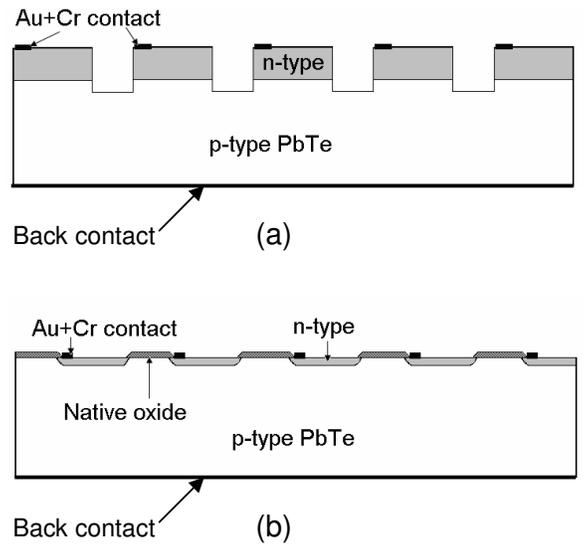

**Figure 1**

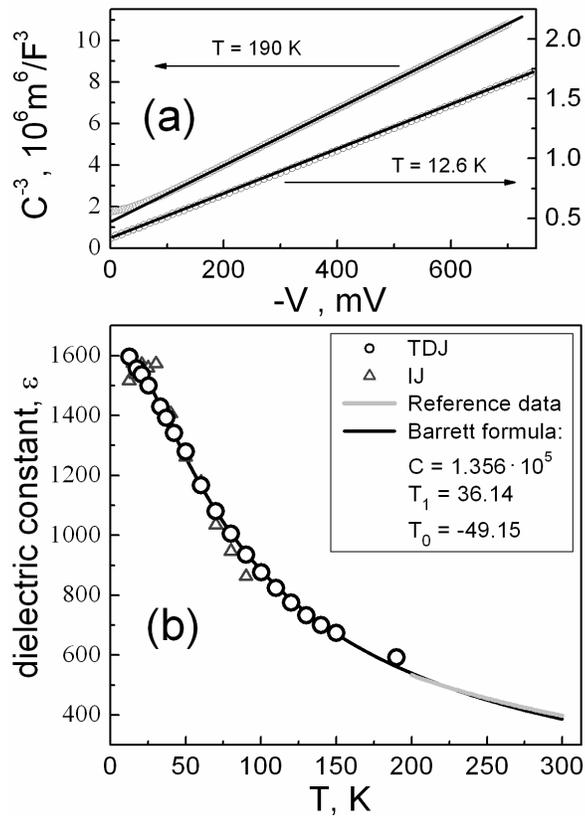

**Fig. 2**

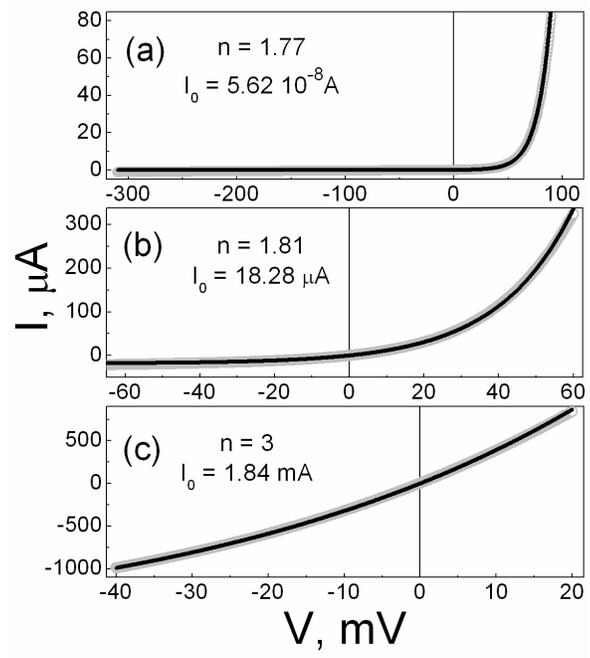

**Fig. 3**

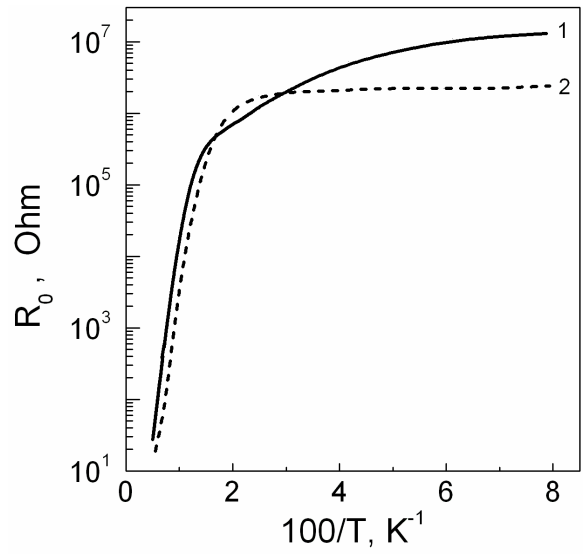

**Fig. 4**

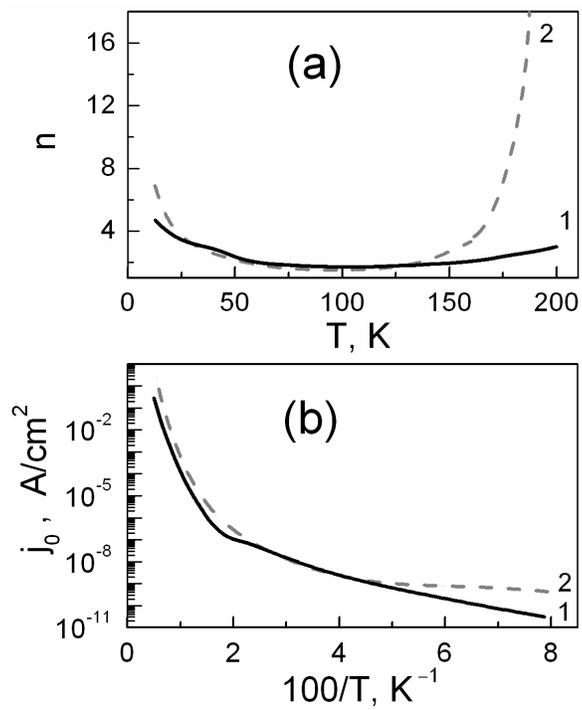

**Fig. 5**